\documentclass[twocolumn,letterpaper,nofootinbib,prl,amsmath]{revtex4}

\usepackage{epsfig}
\usepackage{hyperref}

\newcommand{\lessthansimilarto}{\lower3pt\hbox{$\buildrel{<}\over{\sim}$}}
\newcommand{\greaterthansimilarto}{\lower3pt\hbox{$\buildrel{>}\over{\sim}$}}
\newcommand\LL{{I{\hskip-3pt}L}}

\begin{document}

\title{Modifying the Einstein Equations off the Constraint Hypersurface}

\author{J.~David Brown}
\affiliation{Department of Physics, North Carolina State University,
Raleigh, NC 27695 USA}

\author{Lisa L.~Lowe}
\affiliation{Department of Physics, North Carolina State University,
Raleigh, NC 27695 USA}

\begin{abstract}
A new technique is presented for modifying the Einstein evolution equations off the 
constraint hypersurface. With this approach the evolution equations for the constraints can be 
specified freely. The equations of motion for the gravitational field 
variables are modified by the addition of terms that are linear and nonlocal in the 
constraints. These terms  are obtained from solutions of the linearized Einstein 
constraints. 
\end{abstract}
 
\maketitle

The Einstein equations  separate into a set of evolution equations and a set of constraints. The 
evolution equations are  partial differential equations (PDE's) that determine how the gravitational 
field variables $g_{ab}$ (the spatial metric) and $K_{ab}$ (the extrinsic curvature) 
evolve forward in time. The constraint equations are PDE's that the field 
variables must satisfy at each instant of time. From a Hamiltonian point of view, the evolution 
equations define solution trajectories in phase space with coordinates $g_{ab}$ 
and momenta $K_{ab}$. Physical trajectories are those that 
lie in the constraint  hypersurface, or subspace, of the gravitational phase space.

Einstein's theory of gravity is a  ``first class'' theory, that is, the time derivatives 
of the constraints are linear combinations of the constraints. This property implies that, 
analytically, the constraints will hold at each instant of time if they hold at the initial time. 
However, for numerically generated solutions of the theory
the initial data will not satisfy the constraints precisely and numerical errors will 
kick the phase space trajectory away from the constraint hypersurface. This is a critical problem for 
numerical modeling because the Einstein evolution equations, as they are usually written, 
admit solutions that rapidly diverge away from the constraint 
hypersurface \cite{Kidder:2001tz,Lindblom:2002et}. Any numerical 
scheme that evolves the gravitational field data using the evolution equations in one of their traditional 
forms will eventually fail to produce physically meaningful results. Inevitably the numerical solution will 
choose to follow a trajectory that violates the constraints.

A number of strategies have been devised to address this problem. One approach is to modify 
the theory off the constraint hypersurface by adding linear combinations of constraints 
to the evolution equations 
\cite{Detweiler:1987,Kidder:2001tz,Shinkai:2002yf,Anderson:2003dz,Tiglio:2003cf,Lindblom:2004gd}. 
In this way one hopes to alter the 
solution trajectories so that they are better behaved away from 
the constraint hypersurface. We will use the terminology  ``off--shell'' to refer 
to solution trajectories that lie  off the constraint hypersurface. 

The strategy discussed in this paper 
is of this sort. We add terms proportional to the constraints to the Einstein evolution equations
in such a way that the evolution equations for the constraints can be 
freely specified.  In 
principle we can eliminate all constraint violating modes by demanding, for example, that the 
time derivatives of the constraints should vanish. The price we pay for this degree of control over 
the unphysical, off--shell solutions  is that the terms 
added to the evolution equations are nonlocal. They are determined through the solution of an 
elliptic system of PDE's.

Another strategy for keeping a numerically generated solution from diverging away from  
the constraint hypersurface is constrained evolution. In this scheme the constraints are 
used in place of certain evolution equations to update some of the 
gravitational field variables in time. This approach has worked well for spherically and axisymmetric 
problems 
\cite{Stark:1987qy,Choptuik:1992jv,Abrahams:1992ib,Abrahams:1994ge,Choptuik:2003as}. 
A closely related idea is constraint projection
\cite{Anderson:2003dz,Matzner:2004uu,Holst:2004wt}. With constraint 
projection one evolves the full set of 
field variables using the evolution equations, then periodically (perhaps every timestep) solves the 
constraints to project the solution back to the constraint hypersurface. Both constrained 
evolution and constraint projection require the solution of the constraint equations during the 
course of evolution. For these approaches to be viable, the constraints must be expressed 
as an elliptic system of PDE's. From a 
computational perspective, our strategy   is  closely related to 
constraint projection  since we also solve an elliptic system of PDE's at every (or nearly 
every) timestep. In fact, the PDE's that we solve are 
the linearized  Einstein constraints. 

It will be useful to give an overview of our procedure for modifying the off--shell solutions 
in a  formal, general context. 
Consider a theory, like general relativity, described by a set of first 
class constraints ${\cal C}_A$. Let $\psi_\mu$ denote the basic field variables. 
These variables satisfy first order in time differential equations of motion, 
$\dot \psi_\mu = 
(\dot \psi_\mu)_{\rm old\> rhs}$, where the ``old right--hand sides''
$(\dot \psi_\mu)_{\rm old\> rhs}$ 
are functions of $\psi_\mu$ and their spatial derivatives. 
We have included the descriptor ``old'' 
since we will soon create ``new'' right--hand sides by adding  
functions of the constraints. The evolution equations for the constraints are obtained 
from the evolution equations for the $\psi$'s by differentiating the constraints 
in time. This yields $\dot {\cal C}_A = (\dot{\cal C}_A)_{\rm old\> rhs}$ where 
the right--hand sides (rhs's) are given by 
\begin{equation}\label{Cdoteqn}
  (\dot{\cal C}_A)_{\rm old\> rhs} = \frac{\delta {\cal C}_A}{\delta \psi_\mu}
       (\dot \psi_\mu)_{\rm old\> rhs} \ .
\end{equation}
The expression $\delta {\cal C}_A /\delta \psi_\mu$ is the Fr{\'e}chet derivative of the constraints 
with respect to the field variables \cite{Choquet-Bruhat}. It satisfies 
${\cal C}_A(\psi + \sigma) - {\cal C}_A(\psi) = (\delta{\cal C}_A /\delta \psi_\mu ) \sigma_\mu$, 
in the limit as the norm of $\sigma_\mu$ goes to zero. ($\psi_\mu$ and $\sigma_\mu$ are defined 
as vectors in a suitable Banach space.) If ${\cal C}_A$ depends on spatial derivatives 
of $\psi_\mu$, then the Fr{\'e}chet derivative is a differential operator. 

Now let us split the basic variables $\psi_\mu$ into two sets, $\phi_A$ and $\chi_i$. Note that 
there are as many $\phi$'s as there are constraints. With this splitting, Eq.~(\ref{Cdoteqn}) becomes
\begin{equation}
  (\dot{\cal C}_A)_{\rm old\> rhs} = \frac{\delta {\cal C}_A}{\delta \phi_B}
       (\dot \phi_B)_{\rm old\> rhs} + \frac{\delta {\cal C}_A}{\delta \chi_i}
       (\dot \chi_i)_{\rm old\> rhs}\ .
\end{equation}
Next, we replace the old equations of motion for the $\phi$'s with new equations of motion. 
This leads to new equations of motion for the constraints, 
\begin{equation}
  (\dot{\cal C}_A)_{\rm new\> rhs} = \frac{\delta {\cal C}_A}{\delta \phi_B}
       (\dot \phi_B)_{\rm new\> rhs} + \frac{\delta {\cal C}_A}{\delta \chi_i}
       (\dot \chi_i)_{\rm old\> rhs}\ .
\end{equation}
By subtracting the previous two results, we find
\begin{equation}\label{FormalPhiEqn}
   \Lambda_A = \frac{\partial {\cal C}_A}{\partial \phi_B} \Phi_B 
\end{equation}
where $\Phi_A$ is the difference between the new and old rhs's for the variables $\phi_A$, 
$ \Phi_A \equiv (\dot \phi_A)_{\rm new\> rhs} - 
(\dot \phi_A)_{\rm old\> rhs}$, and $\Lambda_A$ is the difference between 
the new and old rhs's for the constraints, 
$ \Lambda_A \equiv (\dot {\cal C}_A)_{\rm new\> rhs} - 
(\dot {\cal C}_A)_{\rm old\> rhs}$. In terms of the original field 
variables $\psi_\mu$, the new equations of motion are 
\begin{equation}\label{FormalEqnsOfMotion}
   (\dot \psi_\mu)_{\rm new\> rhs} = 
    (\dot \psi_\mu)_{\rm old\> rhs} + 
    \frac{\delta \psi_\mu}{\delta \phi_A} \Phi_A \ .
\end{equation}
Now we turn the reasoning around. We do not actually choose new equations of motion for the $\phi$'s. 
Instead,  we 
specify  new evolution equations for the constraints by freely choosing the expressions 
$(\dot{\cal C}_A)_{\rm new\> rhs}$. 
The functions $\Lambda_A$ are then determined, 
and Eqs.~(\ref{FormalPhiEqn}) are solved for  $\Phi_A$. The new equations of motion for the original 
field variables are given by Eqs.~(\ref{FormalEqnsOfMotion}). 

Because the theory is first class, 
$(\dot{\cal C}_A)_{\rm old\> rhs}$ is a linear combination of constraints. Let 
us  choose  $(\dot{\cal C}_A)_{\rm new\> rhs}$ to be a linear combination 
of constraints as well. Then $\Lambda_A$ is a linear combination of constraints and, according to 
Eq.~(\ref{FormalPhiEqn}), $\Phi_A$ is a (possibly nonlocal) linear combination of constraints. It 
follows that the new 
equations of motion for $\psi_\mu$ differ from the old equations by a linear combination of 
constraints. 

Equations (\ref{FormalPhiEqn}) are the linearized constraints. To be precise, 
consider a field configuration $\psi_\mu$ that does not satisfy the constraints and let $\bar\psi_\mu
= \psi_\mu - \Psi_\mu$. If $\bar\psi_\mu$  satisfies the constraints then to linear order, $\Psi_\mu$
satisfies ${\cal C}_A = (\delta {\cal C}_A/\delta \psi_\mu) \Psi_\mu$. 
This is Eq.~(\ref{FormalPhiEqn}) with $\Psi_\mu = (\delta\psi_\mu/\delta\phi_A) \Phi_A$ and 
${\cal C}_A = \Lambda_A$. 

With the procedure outlined above we can freely specify the rhs's of the constraint evolution equations, 
as long as they are 
linear combination of constraints. We then solve  Eq.~(\ref{FormalPhiEqn}) for 
$\Phi_A$ and modify the equations of motion for the $\psi$'s to Eq.~(\ref{FormalEqnsOfMotion}). 
In this way we leave the equations of 
motion for the  basic field variables unchanged on the constraint hypersurface, 
but we can modify their off--shell form to eliminate the constraint violating modes. 

Let us apply this formalism to general relativity. The basic field variables 
(the $\psi$'s) are the spatial metric $g_{ab}$ and the extrinsic curvature $K_{ab}$. The equations of 
motion as written by York \cite{York:Smarrarticle} are 
$\partial_\perp g_{ab} = (\partial_\perp g_{ab})_{\rm old\> rhs}$ and 
$\partial_\perp K_{ab} = (\partial_\perp K_{ab})_{\rm old\> rhs}$, where
\begin{subequations}\label{YorkEqns}
\begin{eqnarray}
   \left(\partial_\perp g_{ab}\right)_{\rm old\> rhs} & = &  - 2\alpha K_{ab} \ ,\\
   \left(\partial_\perp K_{ab}\right)_{\rm old\> rhs} & = & \alpha (K K_{ab} - 2 K_{ac} K^c_b + R_{ab})
          \nonumber \\
     & &  - D_a D_b\,\alpha \ .\ \ 
\end{eqnarray}
\end{subequations}
Here, $\alpha$ is the lapse function, $D_a$ is the spatial covariant derivative, and $R_{ab}$ is the 
spatial Ricci tensor. The operator $\partial_\perp \equiv \partial_t - {\cal L}_\beta$ 
is the difference between the time derivative and the Lie derivative along the shift vector $\beta^a$. 
This operator plays the role of the ``dot'' in the formal analysis.

The constraints for general relativity are 
\begin{subequations}\label{Constraints}
  \begin{eqnarray}
   {\cal H} & \equiv & K^2 - K_{ab}K^{ab} + R \ ,\\
   {\cal M}_a & \equiv & D_b K^b_a - D_a K \ .
  \end{eqnarray}
\end{subequations}
With the equations of motion (\ref{YorkEqns}), we find 
\begin{subequations}
  \begin{eqnarray}
   (\partial_\perp {\cal H})_{\rm old\> rhs} & = & 2\alpha K {\cal H} - 2\alpha D_a {\cal M}^a 
      - 4{\cal M}^a D_a\alpha \ ,{\qquad}\\
   (\partial_\perp {\cal M}_a)_{\rm old\> rhs} & = & \alpha K {\cal M}_a  - {\cal H} D_a\alpha 
    - \alpha D_a{\cal H}/2 \ .
  \end{eqnarray}
\end{subequations}
for the rhs's of the constraint evolution equations. 

In order to proceed, we must select a subset of the variables $g_{ab}$, $K_{ab}$ to play the role of the 
$\phi$'s. We will use the conformal transverse--traceless decomposition developed by 
Lichnerowicz and York for solving the initial value problem (see, for example, Ref.~\cite{Cook:2000vr}). 
To begin, we split the metric and extrinsic curvature into 
\begin{subequations}
  \begin{eqnarray}
   g_{ab} & = & \varphi^4 {\tilde g}_{ab} \ ,\\
   K_{ab} & = & \varphi^{-2} A_{ab} + \frac{1}{3} \varphi^4 {\tilde g}_{ab} \tau \ ,
  \end{eqnarray}
\end{subequations}
where $\varphi$ is the conformal factor, ${\tilde g}_{ab}$ is the conformal metric, $\tau$ is the 
trace of the extrinsic curvature, and $A_{ab}$ is symmetric and trace free. Note that these definitions 
are invariant under the conformal transformation \cite{York:conformalpaper,Brown:2005aq} 
${\tilde g}_{ab} \to \xi^4 {\tilde g}_{ab}$, 
$\varphi \to \xi^{-1}\varphi$, $A_{ab} \to \xi^{-2}A_{ab}$, $\tau \to \tau$. 

The tensor $A_{ab}$ can be decomposed in terms of a symmetric, transverse, traceless 
tensor $B_{ab}$ and a vector $w_a$,
\begin{equation}\label{Adecomposition}
   A_{ab} = (\tilde\LL w)_{ab} + B_{ab} \ .
\end{equation}
The operator $\tilde \LL$ is defined by $(\tilde\LL w)_{ab} \equiv {\tilde D}_a w_b + 
{\tilde D}_b w_a - 2{\tilde g}_{ab} {\tilde D}_c w^c/3$. It follows that $w_a$ satisfies the elliptic 
equation ${\tilde D}^b(\tilde\LL w)_{ab} = {\tilde D^b}A_{ab}$. The conformal transformation rule
for $B_{ab}$ is $B_{ab} \to \xi^{-2} B_{ab}$ and the rule for $w_a$ is defined by the relation
$({\tilde\LL w})_{ab} \to \xi^{-2} ({\tilde\LL w})_{ab}$. 

Let $\varphi$ and $w_a$ play the role of the $\phi$'s in Eqs.~(\ref{FormalPhiEqn}) and 
(\ref{FormalEqnsOfMotion}). The Fr{\'e}chet derivatives that appear in those equations 
are somewhat tedious but straightforward to compute. If we let $\Phi$ and $W_a$ denote 
the unknowns (the $\Phi_A$'s) in Eq.~(\ref{FormalPhiEqn}), we find the following results:
\begin{subequations}\label{CCeqnswithphi}
\begin{eqnarray}
   \Lambda_0 & = & -8D^2(\Phi/\varphi) - 2K^{ab} (\tilde\LL W)_{ab}/\varphi^2 \nonumber \\
    & & - \left[ 4K^2 - 12K_{ab} K^{ab} + 4R\right] \Phi/\varphi \ , \\
   \Lambda_a & = & D^b\left[ (\tilde\LL W)_{ab} /\varphi^2\right] \nonumber \\
   & & -6\left( D^b K_{ab} 
   - D_aK/3\right)\Phi/\varphi  \ .
\end{eqnarray}
\end{subequations}
Here, $\Lambda_0$ and $\Lambda_a$ are the differences between the new and old rhs's of the 
evolution equations for the constraints:
\begin{subequations}\label{Lambdadefs}
\begin{eqnarray}
   \Lambda_0 & \equiv & \left(\partial_\perp {\cal H} \right)_{\rm new\> rhs} - 
       \left(\partial_\perp {\cal H} \right)_{\rm old\> rhs}  \ ,\\
   \Lambda_a & \equiv & \left(\partial_\perp {\cal M}_a \right)_{\rm new\> rhs} - 
       \left(\partial_\perp {\cal M}_a \right)_{\rm old\> rhs}  \ .
\end{eqnarray}
\end{subequations}
The unknowns in 
Eqs.~(\ref{CCeqnswithphi}) are the differences between the new and old rhs's for 
the $\varphi$ and $w_a$ equations of motion. 
From the formal Eq.~(\ref{FormalEqnsOfMotion}), we find the new equations of motion 
\begin{subequations}\label{NewEOMswithphi}
 \begin{eqnarray}
   \left( \partial_\perp g_{ab} \right)_{\rm new\> rhs} & = & 
     \left( \partial_\perp g_{ab}\right)_{\rm old\> rhs} + 4 g_{ab} \Phi/\varphi \ ,\\
   \left( \partial_\perp K_{ab} \right)_{\rm new\> rhs} & = & 
      \left( \partial_\perp K_{ab} \right)_{\rm old\> rhs}  + (\tilde\LL W)_{ab}/\varphi^2 \nonumber\\
   & & - 2(K_{ab} - K g_{ab}) \Phi/\varphi  
  \end{eqnarray}
\end{subequations}
for the metric and extrinsic curvature.

Let us assume that the new equations of motion for $\varphi$ and $w_a$, like the old, are 
conformally invariant. Then we see that $\Phi$ and $W_a$ inherit the conformal transformation properties 
$\Phi \to \xi^{-1} \Phi$ and $(\tilde\LL W)_{ab} \to \xi^{-2} (\tilde\LL W)_{ab}$. 
It follows that Eqs.~(\ref{CCeqnswithphi}) and (\ref{NewEOMswithphi}) 
are invariant under conformal transformations. 
In other words, these equations hold for any choice of splitting 
of the physical metric $g_{ab}$ into a conformal factor $\varphi^4$ and conformal metric ${\tilde g}_{ab}$. 
For simplicity we can choose the conformal factor 
to be unity, $\varphi = 1$, and the conformal metric to coincide with the physical metric, ${\tilde g}_{ab} 
= g_{ab}$. Then Eqs.~(\ref{CCeqnswithphi}) become 
\begin{subequations}\label{CCeqns}
\begin{eqnarray}
      \Lambda_0 & = & -8D^2\Phi - 2K^{ab} (\LL W)_{ab} \nonumber \\
    & & - \left[ 4K^2 - 12K_{ab} K^{ab} + 4R\right] \Phi \ , \\
   \Lambda_a & = & D^b (\LL W)_{ab} 
    -6\left( D^b K_{ab} 
   - D_aK/3\right)\Phi \ .\qquad
\end{eqnarray}
\end{subequations}
This is an elliptic system of linear PDE's for $\Phi$ and $W_a$.
The new equations of motion (\ref{NewEOMswithphi}) become 
\begin{subequations}\label{NewEOMs}
\begin{eqnarray}
  \left( \partial_\perp g_{ab} \right)_{\rm new\> rhs} & = & 
     \left( \partial_\perp g_{ab}\right)_{\rm old\> rhs} + 4 g_{ab} \Phi \ ,\\
   \left( \partial_\perp K_{ab} \right)_{\rm new\> rhs} & = & 
      \left( \partial_\perp K_{ab} \right)_{\rm old\> rhs}  + (\LL W)_{ab} \nonumber\\
   & & - 2(K_{ab} - K g_{ab}) \Phi \ .
\end{eqnarray}
\end{subequations}
This is our main result. We can now specify new rhs's for the 
Hamiltonian and momentum constraint equations of motion. The $\Lambda$'s are found 
from  Eqs.~(\ref{Lambdadefs}), and Eqs.~(\ref{CCeqns}) 
are solved for $\Phi$ and $W_a$. These results are used in the new equations of motion, 
Eqs.~(\ref{NewEOMs}). Note that these equations contain only the physical metric and 
extrinsic curvature---all references to the conformal transverse--traceless splitting have disappeared. 

What follows is a numerical demonstration that Eqs.~(\ref{CCeqns}), (\ref{NewEOMs}) allow us the freedom 
to prescribe the evolution of the constraints by altering the Einstein equations off--shell.
For this test we use periodic
identification in the $x$, $y$, and $z$ coordinate directions with periods of $2\pi$. Thus, 
for now, we intentionally avoid facing the very important issue of boundary conditions. 
We use initial data that violates the constraints. The spatial metric is 
flat with Cartesian coordinates,  $g_{ab} = \delta_{ab}$. The diagonal 
components of the extrinsic curvature are given by $K_{xx} = K_{yy} = K_{zz} = A/3$. The 
off--diagonal component $K_{xy}$ is 
\begin{equation}
  K_{xy} = \varepsilon_1 \cos^2(z) + \varepsilon_2 \cos(x) \cos(y) \ .
\end{equation}
The components $K_{yz}$ and $K_{zx}$ are obtained from $K_{xy}$ by cyclic permutations of 
$x$, $y$ and $z$. The initial data is evolved with unit lapse $\alpha = 1$ and vanishing 
shift $\beta^a = 0$. For the test results shown here we use the values 
$A = 0.02$, $\varepsilon_1 = 0.01$, and $\varepsilon_2 = 0.0005$.

\begin{figure}[!htb]
\begingroup%
  \makeatletter%
  \newcommand{\GNUPLOTspecial}{%
    \@sanitize\catcode`\%=14\relax\special}%
  \setlength{\unitlength}{0.1bp}%
\begin{picture}(2448,1943)(0,0)%
\special{psfile=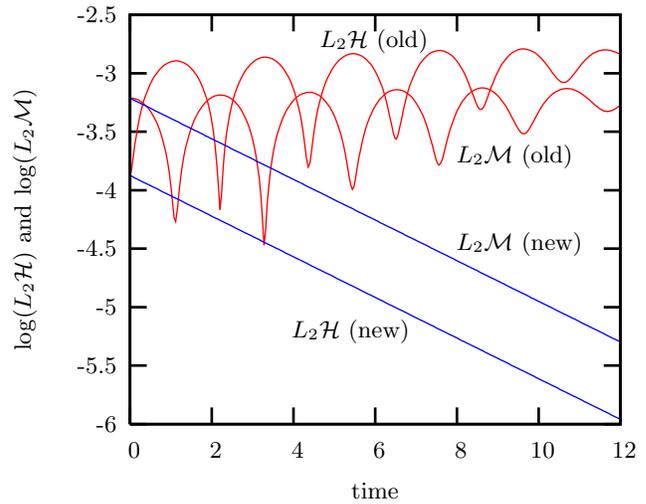 llx=0 lly=0 urx=245 ury=194 rwi=2450}
\put(1680,980){\makebox(0,0)[l]{$L_2{\cal M}$ (new)}}%
\put(1060,650){\makebox(0,0)[l]{$L_2{\cal H}$ (new)}}%
\put(1680,1310){\makebox(0,0)[l]{$L_2{\cal M}$ (old)}}%
\put(1165,1745){\makebox(0,0)[l]{$L_2{\cal H}$ (old)}}%
\put(1374,50){\makebox(0,0){time}}%
\put(100,1072){%
\special{ps: gsave currentpoint currentpoint translate
270 rotate neg exch neg exch translate}%
\makebox(0,0)[b]{\shortstack{$\log(L_2{\cal H})$ and $\log(L_2{\cal M})$}}%
\special{ps: currentpoint grestore moveto}%
}%
\put(2298,200){\makebox(0,0){ 12}}%
\put(1990,200){\makebox(0,0){ 10}}%
\put(1682,200){\makebox(0,0){ 8}}%
\put(1374,200){\makebox(0,0){ 6}}%
\put(1066,200){\makebox(0,0){ 4}}%
\put(758,200){\makebox(0,0){ 2}}%
\put(450,200){\makebox(0,0){ 0}}%
\put(400,1844){\makebox(0,0)[r]{-2.5}}%
\put(400,1623){\makebox(0,0)[r]{-3}}%
\put(400,1403){\makebox(0,0)[r]{-3.5}}%
\put(400,1182){\makebox(0,0)[r]{-4}}%
\put(400,962){\makebox(0,0)[r]{-4.5}}%
\put(400,741){\makebox(0,0)[r]{-5}}%
\put(400,521){\makebox(0,0)[r]{-5.5}}%
\put(400,300){\makebox(0,0)[r]{-6}}%
\end{picture}%
\endgroup
 
\caption{Common logarithm of the $L_2$ norms of the Hamiltonian and momentum constraints, using 
the old and new equations of motion.}
\label{fig1}
\end{figure}

Figure \ref{fig1} shows the common logarithm of the $L_2$ norm of the constraints as  a function of time, 
with and without the off--shell modification terms in Eqs.~(\ref{NewEOMs}). 
For the new constraint equations  we have chosen
\begin{subequations}
\begin{eqnarray}
   (\partial_\perp {\cal H})_{\rm new\> rhs} & = & -0.4{\cal H} - {\cal L}_\beta{\cal H} \ ,\\
   (\partial_\perp {\cal M}_a)_{\rm new\> rhs} & = & -0.4{\cal M}_a - {\cal L}_\beta{\cal M}_a \ ,
\end{eqnarray}
\end{subequations}
so that at each point in space ${\cal H}$ and ${\cal M}_a$ have time dependence $\sim \exp(-0.4 t)$. The 
$L_2$ norms are defined without any factors of $g_{ab}$ or $\sqrt{g}$. That is, we define 
$L_2{\cal H} = \sqrt{\sum {\cal H}^2 / N^3}$ where the sum extends over the $N^3$ grid points. 
Similarly, we define $L_2{\cal M} = \sqrt{\sum {\cal M}_a {\cal M}_a /N^3}$ and include 
a sum over the index $a$.   With these definitions, the only 
time dependence that should appear in the $L_2$ norms is the exponential decay $\sim \exp(-0.4 t)$. 
This is precisely 
what we see in Fig.~(\ref{fig1}) with the new equations of motion. For the old equations of motion,
we see strong oscillations on a short time scale and exponential growth on a longer time scale.

Our numerical code uses pseudospectral collocation \cite{Boyd} with a Fourier basis 
in each of the coordinate directions. Fourth---order Runge--Kutta is used for time stepping. 
The elliptic equations (\ref{CCeqns}) are solved with the iterative method GMRES \cite{Kelley}. 
We use a left preconditioner consisting of the inverse of the diagonal part of the elliptic 
operator. One of the numerical issues that we face is spectral blocking \cite{Boyd}. This 
is the phenomenon in which aliasing causes an unphysical increase in power in the highest 
wave number modes that can be supported on the grid. Filtering can help alleviate 
this problem. For the simulations shown in Fig.~(\ref{fig1}), we use $N = 20$ collocation points 
in each dimension and a filter that sets the two highest frequencies to zero at the end of each timestep. The 
timestep is $0.04$, compared to a light--crossing time of approximately $2\pi$. 

The results displayed in Fig.~(\ref{fig1}) show that we have indeed modified the equations of 
motion off--shell in such a way that unwanted growth in the constraints is eliminated. Ultimately, 
what we would like to show is the ability to prevent constraint growth in the first place. Our 
preliminary attempts to demonstrate this ability have not been completely successful, for reasons that 
we suspect are purely numerical. Although we cannot rule out the possibility that the 
combined system Eqs.~(\ref{CCeqns}), (\ref{NewEOMs}) is mathematically ill--defined in some sense, 
the problems that we have 
encountered appear to be caused by numerical issues. One issue is spectral blocking, mentioned above. 
Another issue is the failure of our elliptic solver to converge to a solution under circumstances 
that we do not yet understand. We suspect 
that a better preconditioner will make our elliptic solver more robust and dependable. 

For the simulation shown in Fig.~(\ref{fig1}), with the new equations of motion, 
the constraints continue to drop exponentially until $t\approx 15$. Beyond
this time the constraints begin to suffer from high wave number variations whose growth counteracts 
the exponential drop of the longer wavelength modes. This breakdown is  sensitive to the 
spatial resolution and amount of filtering and appears to be related to spectral blocking. With the 
old equations of motion, the 
Hamiltonian and momentum constraints continue to grow exponentially until $t\approx 45$. 
At that time $L_2{\cal H}$ has a value of $\sim 10^{-1}$ and the simulation breaks down. Again, this appears 
to be related to spectral blocking. We plan to study these 
issues in more detail and present further numerical tests in a future publication.

\begin{acknowledgments}
We would like to thank Ronald Fulp for helpful discussions. This work was supported by NASA Space Sciences 
grant ATP02--0043--0056 and NSF grant PHY--060042.
\end{acknowledgments}

\bibliography{references}

\end{document}